\documentclass[twocolumn,floatfix,prl,aps,showpacs]{revtex4}
\usepackage{epsfig,graphicx}
\usepackage{flafter}
\usepackage{amsmath}
\usepackage{color}
\usepackage{braket}

\begin{document}

\title{Superstripes and the Excitation Spectrum of a
Spin-Orbit-Coupled Bose-Einstein Condensate}

\author{Yun Li$^1$, Giovanni I. Martone$^1$, Lev P.
Pitaevskii$^{1,2}$, Sandro Stringari$^1$}

\affiliation{$^1$INO-CNR BEC Center and Dipartimento di Fisica,
Universit\`{a} di Trento, I-38123 Povo, Italy\\
$^2$Kapitza Institute for Physical Problems, RAS, Kosygina 2, 119334
Moscow, Russia}

\begin{abstract}
Using Bogoliubov theory we calculate the excitation spectrum of a
spinor Bose-Einstein condensed gas with equal Rashba and Dresselhaus
spin-orbit coupling in the stripe phase. The emergence of a double
gapless band structure is pointed out as a key signature of
Bose-Einstein condensation and of the spontaneous breaking of
translational invariance symmetry. In the long wavelength limit the
lower and upper branches exhibit, respectively, a clear spin and
density nature. For wave vectors close to the first Brillouin zone,
the lower branch acquires an important density character responsible
for the divergent behavior of the structure factor and of the static
response function, reflecting the occurrence of crystalline order.
The sound velocities are calculated as functions of the Raman
coupling for excitations propagating orthogonal and parallel to the
stripes. Our predictions provide new perspectives for the
identification of supersolid phenomena in ultracold atomic gases.

\end{abstract}


\maketitle


The search for supersolidity represents a challenging topic of
research in different areas of condensed matter and atomic physics
(for a recent review see, for example, \cite{Boninsegni2012}).
Supersolidity was first predicted in the pioneering works by Andreev
and Lifschitz \cite{Andreev1969}, Leggett \cite{Leggett1970}, and
Chester \cite{Chester1970}. It is characterized by the coexistence
of two spontaneously broken symmetries. The breaking of gauge
symmetry gives rise to off-diagonal long-range order yielding
superfluidity, while the breaking of translational invariance yields
diagonal long-range order characterizing the crystalline structure.
The First experimental efforts toward the search of supersolidity
were carried out in solid helium \cite{Kim2004}. The strongly
interacting nature of this system makes, however, the effects due to
Bose-Einstein condensation (BEC) extremely small and no conclusive
proof of supersolidity is still available in such a system
\cite{Balibar2010}. More recently, systematic attempts to predict
the occurrence of a supersolid phase have been carried out in atomic
gases with dipolar \cite{Goral2002,Sansone2010,Pollet2010} and soft
core, finite range interactions \cite{Cinti2010,Saccani2011,
Saccani2012,Kunimi2012,Macri2012,Gross}. However, these
configurations have not yet been experimentally realized in the
quantum degenerate phase required to observe the new effects.

The recent realization of spinor BECs with spin-orbit coupling
\cite{Lin2009_PRL, Lin2009_Nature, Lin2011_NatPhy, Lin2011} is
opening new perspectives in the field. In systems with equal Rashba
and Dresselhaus couplings and for small values of the Raman
coupling, theory in fact predicts the occurrence of a stripe phase
where translational invariance is spontaneously broken
\cite{Ho2011,Li2012_PRL,note_rashba}. Actually these systems are
periodic only in one direction and can be considered as superfluid
nematic liquid crystals. Experiments are already available in the
relevant range of parameters, but no direct evidence of the density
modulations is still available, due to the smallness of the contrast
and the microscopic distance separating consecutive stripes. A phase
transition has been nevertheless detected \cite{Lin2011} at values
of the Raman coupling below which theory predicts the occurrence of
the stripe phase.

The purpose of this work is to show that the excitation
spectrum of the gas in the stripe phase exhibits typical supersolid
features, like the occurrence of two gapless bands and the divergent
behavior of the static structure factor for wave vectors approaching
the boundary of the Brillouin zone. The excitation spectrum is
measurable in Bragg spectroscopy experiments, so the experimental
characterization of the new phase should not represent a major
difficulty.

Spin-orbit-coupled BECs can be described using the mean-field
Gross-Pitaevskii picture. The interaction is zero ranged and is
characterized by the values of the scattering lengths associated
with the two hyperfine states involved in the Raman process (we
limit here the discussion to spinor Bose gases). This differs from
the case of other systems, like dipolar gases, where the origin of
the supersolid phase is associated with the finite range of the
force \cite{Gross}. The validity of the Gross-Pitaevskii approach
can be tested {\it a posteriori} by evaluating the quantum depletion
of the condensate.

We use the single-particle Hamiltonian ($\hbar=m=1$)
\begin{equation}
h_0= \frac{1}{2}\left[\left(p_x-k_0 \sigma_z\right)^2+
p_\perp^2\right] + \frac{\Omega}{2} \sigma_x \,, \label{eq:h0}
\end{equation}
accounting for the effect of two counterpropagating and polarized
laser fields, where $k_0$ is fixed by the momentum transfer of the
two lasers, while $\Omega$ is the Raman coupling, accounting for the
intensity of the laser beams causing the transition between the two
spin states. The occurrence of the term $\delta\sigma_z/2$ has been
ignored in $h_0$, since we will consider situations where the
effective magnetic field $\delta$ is zero (experimentally this can
be achieved with a proper detuning of the two laser fields).
Hamiltonian (\ref{eq:h0}) can be formally derived by applying a
unitary transformation to the Hamiltonian in the laboratory frame
describing the system in the presence of two detuned, spin-polarized
laser fields \cite{Martone2012}. The unitary transformation consists
of a local rotation in spin space around the $z$ axis, causing the
appearance of the spin-orbit term proportional to $p_x \sigma_z$.

Remarkable properties of this Hamiltonian are its translational
invariance and, for $\Omega<2k^2_0$, the occurrence of a
double-minimum structure in the single-particle energy at momenta
$p_x=\pm k_1$ with $k_1=k_0\sqrt{1-\Omega^2/4k^4_0}$, capable to
host a BEC. This structure is at the origin of new intriguing
features, like the existence of a spin-polarized plane-wave phase
and of an unpolarized stripe phase \cite{Ho2011,Li2012_PRL} at even
smaller values of $\Omega$, resulting from the spontaneous breaking
of translational symmetry. From general arguments one expects that
the spontaneous breaking of this continuous symmetry is at the
origin of a new gapless Goldstone mode.

\begin{figure}
\centering
\includegraphics[scale=0.42]{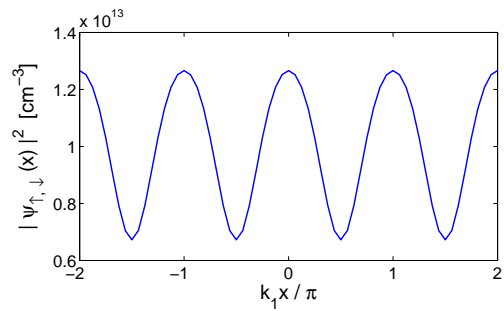}
\caption{Density profile along the $x$ direction. The parameters are
$\Omega/k_0^2=1.0$, $G_1/k_0^2=0.3$, and $ G_2/k_0^2= 0.08$.}
\label{fig:dens_prof}
\end{figure}

The stripe phase arises due to the competition between the
density and spin-density interaction terms in the mean-field
Hamiltonian
\begin{equation}
H_{\text{int}}= \frac{1}{4} \int d^3r \left[
\left(g+g_{\uparrow\downarrow} \right) n(\mathbf{r})^2 +
\left(g-g_{\uparrow\downarrow}\right) s(\mathbf{r})^2 \right]
\label{eq:E_int}
\end{equation}
where $n(\mathbf{r})= n_\uparrow(\mathbf{r})+ n_\downarrow(
\mathbf{r})$ and $s(\mathbf{r})=n_\uparrow(\mathbf{r})-
n_\downarrow(\mathbf{r})$ correspond to the total and spin
densities. In Eq.~(\ref{eq:E_int}) we have assumed equal
intraspecies interactions $g_{\uparrow\uparrow}=
g_{\downarrow\downarrow} \equiv g$ \cite{note_scattering} with
$g_{\alpha\beta}$ ($\alpha,\,\beta =\, \uparrow,\,\downarrow$) being
the coupling constants in the different spin channels. The stripe
phase emerges only for $g_{\uparrow\downarrow}< g$, a condition
yielding an unpolarized uniform ground state in the absence of the
Raman coupling $\Omega$. It is associated with the macroscopic
occupation of a single-particle spinor state of the form
\begin{equation}
\begin{pmatrix} \psi_{0\uparrow}\\ \psi_{0\downarrow} \end{pmatrix} =
\sum_{\bar{K}} \begin{pmatrix} \;\;\, a_{-k_1+\bar{K}} \, \\
-b_{-k_1+\bar{K}}\, \end{pmatrix} e^{i \left(\bar{K} - k_1 \right)
x} \label{eq:ground_psi}
\end{equation}
where $k_1 = \pi/d$ is related to the period $d$ of the stripes,
$\bar{K}=2 n k_1$, with $n=0, \,\pm 1, \,\ldots\,$, are the
reciprocal lattice vectors while $a_{-k_1+\bar{K}}$ and
$b_{-k_1+\bar{K}}$ are expansion coefficients to be determined,
together with the value of $k_1$, by a procedure of energy
minimization, including the single-particle (\ref{eq:h0}) and the
interaction (\ref{eq:E_int}) terms in the Hamiltonian. In the stripe
phase, energy minimization gives rise to the presence of terms with
opposite phase ($e^{\pm ik_1x}, \,e^{\pm 3ik_1x},\, \ldots\,$),
responsible for the density modulations and characterized by the
symmetry condition $a_{-k_1+\bar{K}}=b^*_{k_1-\bar{K}}$, causing the
vanishing of the spin polarization. The stripe phase is favored at
small values of the Raman coupling. In the limit of weak
interactions, defined by the condition $G_1, \, G_2 \ll k^2_0$ where
$G_1 = \bar{n}\left(g+ g_{\uparrow\downarrow} \right)/4$ and $G_2 =
\bar{n}\left(g- g_{\uparrow\downarrow} \right)/4$ with $\bar{n}$
being the average density, the critical value for the Raman
frequency is given by $\Omega_{\text{cr}} = 2 k_0^2 \sqrt{2\gamma
/(1+2\gamma)}$ with $\gamma = G_2/G_1$ independent of the density,
and the stripe phase is ensured for values of $\Omega$ smaller than
$\Omega_{\text{cr}}$. In the available experiments with $^{87}$Rb
atoms \cite{Lin2011, Zhang2012}, the value of $G_2$ (and hence
$\Omega_{\text{cr}}$) is very small. To enlarge the range of values
of $\Omega$ compatible with the stripe phase, it is useful to
increase $G_2$ as much as possible. This is crucial to produce a
significant contrast in the density profile which is proportional to
$\Omega/k^2_0$ \cite{Li2012_PRL}. In the following we will use the
values $G_1/k_0^2=0.3$ and $G_2/k_0^2 = 0.08$ yielding
$\Omega_{\text{cr}} /k_0^2 \simeq 1.3$. In Fig.~\ref{fig:dens_prof}
we show the ground state density profile calculated at
$\Omega/k^2_0=1.0$. The other quantum phases predicted by theory at
larger values of $\Omega$ are the plane-wave and the zero momentum
phases. In these phases the sum (\ref{eq:ground_psi}) contains only
the term $e^{ik_1x}$ (or $e^{-ik_1x}$) with $k_1\ne 0$ in the
plane-wave phase and $k_1=0$ in the zero momentum one.

To evaluate the elementary excitations we apply Bogoliubov theory by
writing the deviations of the order parameter with respect to
equilibrium as
\begin{equation}
\begin{pmatrix} \psi_\uparrow\\ \psi_\downarrow \end{pmatrix}
= e^{-i\mu t}\left[\begin{pmatrix} \psi_{0\uparrow} \\
\psi_{0\downarrow}\end{pmatrix} + \begin{pmatrix}
u_{\uparrow}(\mathbf{r}) \\
u_{\downarrow}(\mathbf{r})\end{pmatrix}
e^{-i\omega t} + \begin{pmatrix} v^\ast_{\uparrow}(\mathbf{r})\\
v^\ast_{\downarrow}(\mathbf{r}) \end{pmatrix} e^{i\omega t}\right]
\end{equation}
and solving the corresponding linearized time-dependent
Gross-Pitaevskii equations. The equations are conveniently solved by
expanding $u_{\uparrow,\, \downarrow}(\mathbf{r})$ and
$v_{\uparrow,\, \downarrow}(\mathbf{r})$ in the Bloch form in terms
of the reciprocal lattice vectors:
\begin{eqnarray}
&& u_{\mathbf{q}\,\uparrow,\,\downarrow}(\mathbf{r}) =  e^{-ik_1 x}
\sum_{\bar{K}} U_{\mathbf{q}\,\uparrow,\,\downarrow\,\bar{K}}
\,e^{i\mathbf{q}\cdot \mathbf{r}+i\bar{K}x}\\
&&v_{\mathbf{q}\,\uparrow,\,\downarrow}(\mathbf{r}) =  e^{ik_1 x}
\sum_{\bar{K}} V_{\mathbf{q}\,\uparrow,\,\downarrow\,\bar{K}}
\,e^{i\mathbf{q}\cdot \mathbf{r}-i\bar{K}x}
\end{eqnarray}
where $\mathbf{q}$ is the wave vector of the excitation. The same
ansatz can be used to calculate the density and spin-density
dynamic response function, by adding to the Hamiltonian a
perturbation proportional to $e^{i(\mathbf{q} \cdot
\mathbf{r}-\omega t) +\eta t}$ and $\sigma_z e^{i(\mathbf{q} \cdot
\mathbf{r}-\omega t) +\eta t}$ with $\eta \to 0^+$, respectively.

\begin{figure}[t]
\centering
\includegraphics[scale=0.42]{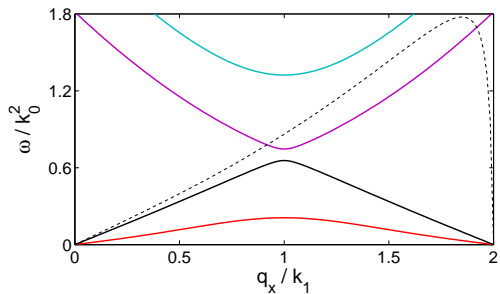}
\caption{(color online). Lowest four excitation bands along the $x$
direction. The parameters are the same as in
Fig.~\ref{fig:dens_prof}. The thin dotted line corresponds to the
Feynman relation $\omega = q_x^2/2 S(q_x)$.} \label{fig:spectrum}
\end{figure}

The excitation spectrum predicted by the Hamiltonian
(\ref{eq:h0})-(\ref{eq:E_int}) has been already calculated in both
the plane-wave phase and the zero momentum phases \cite{Martone2012}
where, despite the spinor nature of the system, only one gapless
branch is predicted as a consequence of the presence of the Raman
coupling $\Omega$. A peculiar feature exhibited by the plane-wave
phase is the emergence of a rotonic structure whose gap becomes
smaller and smaller as one decreases the value of $\Omega$,
providing the onset of the transition to the stripe phase.

\begin{figure}[t]
\centering
\includegraphics[scale=0.42]{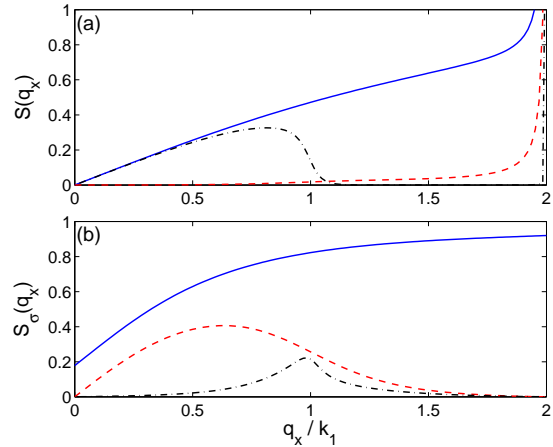}
\caption{(color online). Density (a) and spin-density (b) static
structure factor as a function of $q_x$ (blue solid line). The
contributions of the first (red dashed line) and second (black
dash-dotted line) bands are also shown. The parameters are the same
as in Fig.~\ref{fig:dens_prof}.} \label{fig:static_struc}
\end{figure}

The results for the dispersion law of the elementary excitations in
the stripe phase are reported in Fig.~\ref{fig:spectrum} for the
same parameters used in Fig.~\ref{fig:dens_prof}. We have considered
excitations propagating in the $x$ direction orthogonal to the
stripes and labeled with the wave vector $q_x$. A peculiar feature,
distinguishing the stripe phase from the other uniform phases, is
the occurrence of two gapless bands. At small $q_x$ we find that the
lower branch is basically a spin excitation, while the upper branch
is a density mode, as clearly revealed by
Fig.~\ref{fig:static_struc} (a) where we show the contributions of
the two gapless branches to the static structure factor $S(q_x) =
N^{-1}\sum_\ell| \langle0|\rho_{q_x}| \ell\rangle|^2$ where $\ell$
is the band index and $\rho_{q_x}=\sum_i e^{iq_x x_i}$ is the
density operator. The density nature of the upper branch, at small
$q_x$, is further confirmed by the comparison with the Feynman
relation $\omega= q_x^2/2 S(q_x)$ (see Fig.~\ref{fig:spectrum}). A
two-photon Bragg scattering experiment with laser frequencies far
from resonance, being sensitive to the density response, will
consequently excite only the upper branch at small $q_x$. Bragg
scattering experiments actually measure the imaginary part of the
response function, a quantity which, at enough low temperature, can
be identified with the $T=0$ value of the dynamic structure factor
$S(q_x,\omega)=\sum_\ell| \langle0|\rho_{q_x}| \ell\rangle|^2
\delta(\omega - \omega_{\ell 0})$, where $\omega_{\ell 0}$ is the
excitation frequency of the $\ell\,$th state \cite{bookLevSandro}.
The spin nature of the lower branch is clearly revealed by
Fig.~\ref{fig:static_struc} (b) where we report the contributions
arising from the two gapless branches to the spin static structure
factor $S_\sigma(q_x)=N^{-1} \sum_\ell |\langle0|s_{q_x}|
\ell\rangle|^2$, where $s_{q_x} =\sum_i \sigma_{zi} e^{iq_x x_i}$ is
the spin-density operator. Notice that, differently from $S(q_x)$,
the total spin structure factor does not vanish as $q_x\to 0$, being
affected by the higher energy bands as a consequence of the Raman
term in Hamiltonian (\ref{eq:h0}). The lower branch exhibits a
hybrid nature and, when approaching the Brillouin wave vector
$q_B=2k_1$, it is responsible for the divergent behavior of the
density static structure factor [see Fig.~\ref{fig:static_struc}
(a)], a typical feature exhibited by crystals.

\begin{figure}[t]
\centering
\includegraphics[scale=0.42]{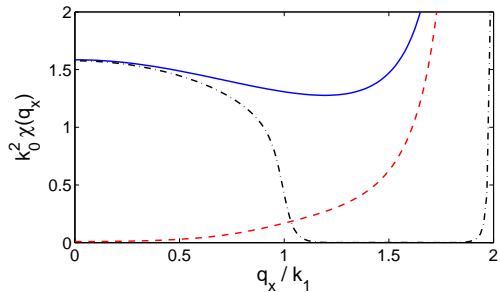}
\caption{(color online). Static response as a function of $q_x$
(blue solid line). The contributions of the first (red dashed line)
and second (black dash-dotted line) bands are also shown. The
parameters are the same as in Fig.~\ref{fig:dens_prof}. }
\label{fig:dens_chi}
\end{figure}

It is worth pointing out that the occurrence of two gapless
excitations is not by itself a signature of supersolidity and is
exhibited also by uniform mixtures of BECs without spin-orbit and
Raman couplings \cite{bookPethick} as well as by the plane-wave
phase of the Rashba Hamiltonian with $SU(2)$ invariant interactions
($G_2=0$) \cite{Barnett2012}. Only the occurrence of a band
structure, characterized by the vanishing of the excitation energy
and by the divergent behavior of the structure factor at the
Brillouin wave vector, can be considered an unambiguous evidence of
the density modulations characterizing the stripe phase. The
divergent behavior near the Brillouin zone is even more pronounced
(see Fig.~\ref{fig:dens_chi}) if one investigates the static
response function $\chi(q_x) = 2N^{-1}\sum_\ell |\langle0|
\rho_{q_x}|\ell\rangle|^2/\omega_{\ell 0}$, proportional to the
inverse energy weighted moment of the dynamic structure factor. The
divergent behaviors of $S(q_x)$ and $\chi(q_x)$ can be rigorously
proven using the Bogoliubov \cite{Bogoliubov1962} and the
Uncertainty Principle \cite{Pitaevskii1991} inequalities applied to
systems with spontaneously broken continuous symmetries. These
inequalities are based, respectively, on the relationships
$m_{-1}(F)m_1(G) \ge |\langle\,[F,G]\,\rangle|^2$ and $m_0(F)m_0(G)
\ge |\langle\, [F,G]\,\rangle|^2$ involving the $p\,$th moments
$m_p(\mathcal{O})=\sum_\ell \left(|\langle0|\mathcal{O}|
\ell\rangle|^2+|\langle 0|\mathcal{O}^\dag|\ell\rangle|^2
\right)\omega_{\ell 0}^p$ of the $\ell\,$th strengths of the
operators $F=\sum_j \,e^{iq_xx_j}$ and $G = \sum_j(p_{xj}\,
e^{-i(q_x -q_B)x_j} + \textrm{H.c.})/2$. The commutator
$\langle\,[F,G]\,\rangle =q_x N\langle e^{iq_Bx} \rangle$, entering
the right-hand side of the inequalities, coincides with the relevant
crystalline order parameter and is proportional to the density
modulations of the stripes \cite{note_Rb}. The moments $m_{-1}(F)$
and $m_0(F)$ are instead proportional to the static response
$\chi(q_x)$ and to the static structure factor $S(q_x)$,
respectively. It is not difficult to show that the moments $m_1(G)$
and $m_0(G)$ are proportional, respectively, to $(q_x-q_B)^2$ and to
$|q_x-q_B|$ as $q_x \to q_B$ due to the translational invariance of
the Hamiltonian. This causes the divergent behaviors $S(q_x)\propto
1/|q_x-q_B|$ and $\chi(q_x)\propto 1/(q_x-q_B)^2$ with a weight
factor proportional to the square of the order parameter
\cite{note_Rb}.

\begin{figure}[t]
\centering
\includegraphics[scale=0.42]{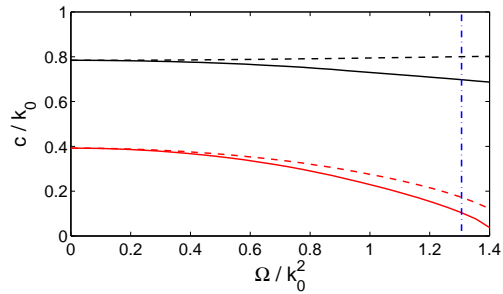}
\caption{(color online). Sound velocities in the first (red) and
second (black) bands along the $x$ ($c_x$, solid lines) and
transverse ($c_{\perp}$, dashed lines) directions as a function of
$\Omega$. The blue dash-dotted line represents the transition from
the stripe phase to the plane-wave phase. The values of the
parameters $G_1/k^2_0$ and $G_2/k^2_0$ are the same as in
Fig.~\ref{fig:dens_prof}.} \label{fig:sound_velo}
\end{figure}

In addition to the excitations propagating along $x$ (longitudinal
modes), another class of Bogoliubov modes is predicted in the
transverse directions, parallel to the stripes. These modes are
excited by the density and spin-density operators $\sum_i e^{iq_y
y_i}$ and $\sum_i\sigma_{zi} e^{iq_y y_i}$. Remarkably, also in the
transverse channel Bogoliubov theory predicts the occurrence of two
gapless spectra. Similarly to the longitudinal channel, at small
$q_y$ the lowest and upper branches have a spin and density
character, respectively.

In Fig.~\ref{fig:sound_velo} we compare the sound velocities of the
two gapless branches in the longitudinal ($c_x$) and transverse
($c_\perp$) directions. We find that $c_x$ is always smaller than
$c_\perp$, reflecting the inertia of the flow caused by the presence
of the stripes. The value of $c_\perp$ in the second band (second
sound) is well reproduced by the Bogoliubov expression $\sqrt{2G_1}$
(equal to $0.78 \,k_0$ in our case) for the sound velocity. Notice
that the spin sound velocity becomes lower and lower as the Raman
frequency increases, approaching the transition to the plane-wave
phase. The Bogoliubov solutions in the stripe phase exist also for
values of $\Omega$ larger than the critical value $\Omega_{cr}= 1.3
\,k_0^2$, due to the first-order nature of the transition.

We have finally checked that the quantum depletion of the
condensate, due to the fluctuations associated with the Bogoliubov
solutions, is always small, thereby confirming the validity of the
mean-field approach.

In conclusion we have shown that the excitation spectrum in the
stripe phase of a spin-orbit-coupled BEC exhibits a double gapless
band structure, typical of supersolids. We predict that at small
wave vectors the lower and upper branches have, respectively, a spin
and density nature. The lower branch, whose gapless nature is due to
the breaking of translational symmetry, is responsible for the
divergent behavior of the static structure factor as the wave vector
approaches the border of the Brillouin zone. The experimental
verification of the new dynamic features predicted in this Letter is
expected to provide a significant advance in our understanding of
systems exhibiting simultaneously off-diagonal and diagonal
long-range order.

\acknowledgments

Useful discussions with G. Shlyapnikov are acknowledged. S.S. likes
to thank the hospitality of the Kavli Institute for Theoretical
Physics. This work has been supported by ERC through the QGBE grant
and by Provincia Autonoma di Trento.


\begin{thebibliography}{99}
\bibitem{Gross}
The possible occurrence of density modulations in the ground state
of a Bose gas interacting with soft core, finite range interactions
was first predicted by Gross \cite{Gross57}.
\bibitem{note_Rb}
This is the main reason why it is useful to work with large values
of the spin interaction parameter $G_2$, allowing for large values
of the Raman coupling $\Omega$ and hence of the crystalline order
parameter. For $^{87}$Rb the value of $G_2$ is small and the
divergency effect in $S(q_x)$ is weak. In this case, the sound
velocity of the lowest band is small and the dispersion practically
exhibits a $q^2$-like behavior at small $q$.
\end{thebibliography}
\end{document}